\documentclass[iop,apj,tighten]{emulateapj}
\pdfoutput=1 
\usepackage{amsmath,amstext}
\usepackage[breaklinks,colorlinks,urlcolor=blue,citecolor=blue,linkcolor=magenta]{hyperref} 
\usepackage[all]{hypcap} 
\usepackage[normalem]{ulem}

\usepackage{float}
\floatstyle{plaintop}
\restylefloat{table}

\usepackage{natbib}
\def\sersic     {S\'ersic}

\shorttitle{Color-mass and morphology at $z\sim 1$}
\shortauthors{Powell et al.}

\begin{document}

\title{Morphology and the Color-Mass Diagram as Clues to Galaxy Evolution at $z\sim 1$}
\author{Meredith C. Powell\altaffilmark{1}, C. Megan Urry\altaffilmark{1}, Carolin N. Cardamone\altaffilmark{2}, Brooke D. Simmons\altaffilmark{3}$^{\dagger}$, Kevin Schawinski\altaffilmark{4}, Sydney Young\altaffilmark{1}, Mari Kawakatsu\altaffilmark{1}}
\affil{$^1$ Yale Center for Astronomy and Astrophysics, and Physics Department, Yale University, PO Box 2018120, New Haven, CT 06520-8120}
\affil{$^2$ Wheelock College, Boston, MA 02215}
\affil{$^3$ Center for Astrophysics and Space Sciences, University of California, San Diego, La Jolla, CA 92093, $^{\dagger}$Einstein Fellow}
\affil{$^4$ Institute for Astronomy, Department of Physics, ETH Zurich, Zurich 8093, Switzerland}

\begin{abstract}
We study the significance of mergers in the quenching of star formation in galaxies at $z \sim 1$, by examining their color-mass distributions for different morphology types. We perform 2-dimensional light profile fits to GOODS $iz$ images of $\sim$5,000 galaxies and X-ray selected active galactic nucleus (AGN) hosts in the CANDELS/GOODS-north and south fields in the redshift range $0.7 < z < 1.3$. Distinguishing between bulge-dominated and disk-dominated morphologies, we find that disks and spheroids have distinct color-mass distributions, in agreement with studies at $z\sim 0$. The smooth distribution across colors for the disk galaxies corresponds to a slow exhaustion of gas, with no fast quenching event. Meanwhile, blue spheroids most likely come from major mergers of star-forming disk galaxies, and the dearth of spheroids at intermediate green colors is suggestive of rapid quenching. The distribution of moderate luminosity X-ray AGN hosts is even across colors, in contrast, and we find similar numbers and distributions among the two morphology types with no apparent dependence on Eddington ratio. The high fraction of bulge-dominated galaxies that host an AGN in the blue cloud and green valley is consistent with the scenario in which the AGN is triggered after a major merger, and the host galaxy then quickly evolves into the green valley. This suggests AGN feedback may play a role in the quenching of star formation in the minority of galaxies that undergo major mergers.

\end{abstract}

\maketitle

\section{Introduction}

The distribution of galaxies on color-mass diagrams can be used to study how galaxies evolve. The stellar mass of a galaxy tracks its hierarchical growth via mergers, as well as its integrated star formation, and the dust-corrected rest-frame colors reflect its current state of star formation. Both local and high redshift galaxies show bimodal color-mass distributions \citep[e.g.,][]{Strateva:2001aa, Baldry:2004aa, Baldry:2006aa, Brammer:2009aa}, with a `blue cloud' of star forming galaxies and a `red sequence' of more massive, quiescent galaxies which build up over cosmic time. The relative lack of galaxies in the `green valley' between these two populations is generally interpreted as evidence of rapid quenching, after which galaxies migrate quickly to the red sequence as the stellar population ages \citep{Bell:aa, Faber:2007aa}. \par

Feedback from active galactic nuclei (AGNs) is a possible quenching mechanism frequently assumed in galaxy evolution models. The tight correlations between supermassive black hole mass and galaxy  properties suggest there is some kind of co-evolution between them, with mergers possibly being the common cause of both AGN accretion and bulge mass buildup \citep{Hopkins:2007aa}. The enormous amount of energy emitted during the AGN phase may couple to the surrounding gas and heat or expel it from the galaxy, thus suppressing further star formation. In addition, AGN feedback is regularly invoked in cosmological simulations to recover the observed local luminosity function, the bimodality, and the number density of massive galaxies \citep[e.g.,][]{Somerville:2008aa}, although these simulations have not yet converged on the strength of the effect. \par

Direct evidence of AGN feedback remains elusive, and recent studies of the relationship between AGN activity and star formation have arrived at different conclusions \citep[e.g.,][]{Hickox:2013aa, Schawinski:2014aa}. Moreover, while AGN hosts are found to reside predominantly in the green valley, several recent studies have suggested that mergers do not play a dominant role in triggering most AGN \citep{Grogin:2005aa, Gabor:2009aa, Cisternas:2011aa, Kocevski:2011aa, Schawinski:2012aa, Treister:2012aa, Simmons:2012aa, Rosario:2015aa}.

Galaxy morphology holds clues to past merger history, as bulges and elliptical galaxies typically form from merger events. It can therefore can be used to distinguish between galaxies on different evolutionary tracks if mergers are significant in evolving the galaxy. \citet{Schawinski:2014aa} showed that local disk-dominated galaxies do not show signs of fast quenching; rather, they evolve slowly until the halo mass reaches a critical point ($M \sim 10^{12} M_{\odot}$), after which they redden more dramatically. Meanwhile the early-type galaxies, assumed to have undergone a major merger, migrate rapidly through the green valley to the red sequence, as per the standard paradigm. \citet{Smethurst:2015aa} built on this using Bayesian inference to create a statistically rigorous approach to finding the star formation histories of the same galaxy sample, and found similar results. However these studies address only local galaxies, for which star formation rates and black hole accretion rates are very low, and thus the conclusions may have little to do with galaxy evolution at the epoch of peak star formation and black hole growth at $z\sim 1-3$. \par

In this paper, we examine the morphology-dependent evolution of higher redshift galaxies toward the peak of galaxy evolution, at $z\sim 1$. We take advantage of deep, multi-wavelength {\it Hubble Space Telescope (HST)} imaging, for which $z\sim 1$ is comparable in resolution and sensitivity to Sloan Digital Sky Survey (SDSS) imaging of $z\sim 0.03$ galaxies, and use galaxy morphologies to indicate which galaxies have undergone a recent major merger. We then compare to matched samples of \textbf{X-ray} AGNs. This paper is organized as follows. We describe the {\it HST} data and our AGN selection in Section 2; Section 3 describes the morphological analysis and sample cuts; results are presented and discussed in Section 4 for both inactive and active galaxy populations; and we present main conclusions in Section 5. We assume a standard $\Lambda$CDM Cosmology ($\Omega_{m} = 0.3$ $\Omega_{\Lambda} = 0.7$ and $H_{0} = 70$ km s$^{-1}$ Mpc$^{-1}$). All magnitudes are in the AB system.

\section{Data}

\begin{figure}[h]
\centering
\includegraphics[width=.4\textwidth]{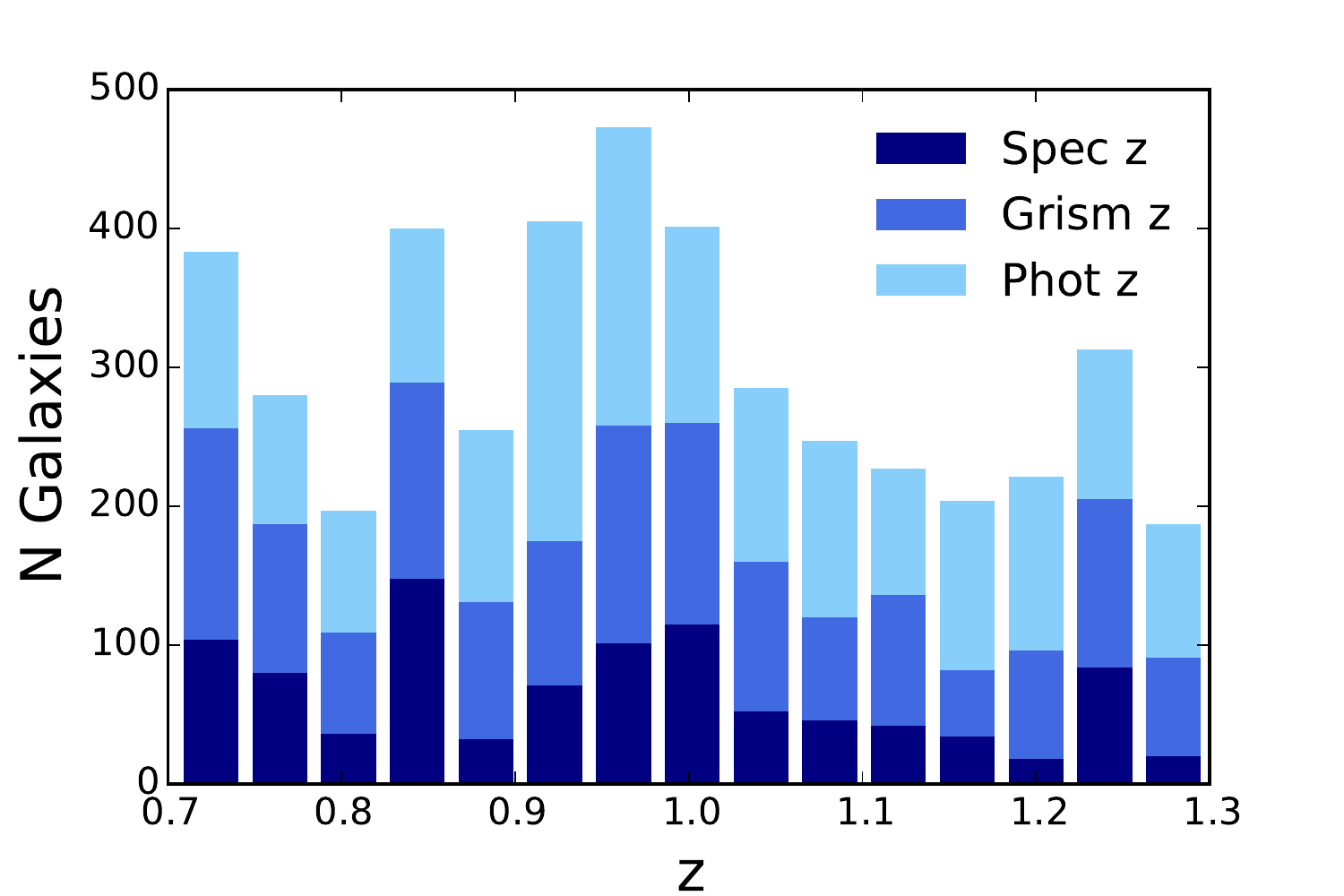}
\caption{Redshift distribution of $z\sim 1$ galaxies. The colors indicate whether the redshifts are based on moderate resolution spectroscopy (dark blue), lower resolution grism spectroscopy (blue) or photometry (light blue). 57$\%$ of redshifts used in this study are spectroscopic and $43\%$ are photometric.}
\label{fig:zhist}
\end{figure}

For this analysis we used {\it HST} ACS data from the north and south fields of the Great Observatories Origins Deep Survey (GOODS; \citealp{Giavalisco:2004aa}). Optical observations were taken in F435W {\it (B)}, F660W {\it(V)}, F775W {\it (i)}, and F850LP {\it (z)} bands, each with a resolution of $0^{\prime \prime}.03$ per pixel. We used the $i$ and $z$ band images for $z \sim 1$ galaxies in order to fit their morphologies in the rest-frame {\it B} band. The magnitude limits for extended sources are $i < 27.7$ mag and $z < 27.3$ mag. We used galaxies in the redshift range $0.7<z<1.3$ that were in the footprint of the Cosmic Assembly Near-IR Deep Extragalactic Legacy Survey (CANDELS; \citealp{Grogin:2011aa}; \citealp{Koekemoer:2011aa}), so that deep near-infrared WFC3/IR data from HST as well as an abundance of other multi-wavelength data were available. The depth and multi-wavelength coverage of the CANDELS fields make them ideal for investigating galaxies and AGN hosts at $z\sim 1$.\par

We used catalogs from the 3D-{\it HST} spectroscopic survey, which provides grism spectroscopy in addition to multiwavelength photometric data over the five CANDELS fields \citep{Brammer:2012aa, Momcheva:2015aa}; ground-based spectroscopy exists for a fraction of the galaxies. The coverage of the two GOODS fields in this survey are 116 arcmin$^{2}$ and 147 arcmin$^{2}$ for north and south, respectively.  For the objects with photometry only, \citet{Skelton:2014} derived photometric redshifts by fitting each object's spectral energy distribution (SED) with a linear combination of seven EAZY galaxy templates  \citep{Brammer:2008aa}, obtaining $\Delta z/(1+z) = 0.026$ and $9.7\%$ outliers for GOODS-N and 
$\Delta z/(1+z) = 0.01$ and $5.4\%$ outliers for GOODS-N. For objects with grism data, \citet{Momcheva:2015aa} fit the two-dimensional (2D) spectra and photometry simultaneously to produce redshifts with $\Delta z/(1+z) = 0.003$. The authors then obtained the rest-frame colors by using the best-fitting redshift with the EAZY templates, and derived the stellar masses and extinction values from SED fits to stellar population models with the FAST code \citep{Kriek:2009aa}, as described in \citet{Skelton:2014} and \citet{Brammer:2012aa}. Their star formation rates were obtained by including UV light from massive stars re-radiated in the FIR, calculated in \citet{Whitaker:2014aa}. More details on the 3D-{\it HST} survey can be found in \citet{Momcheva:2015aa} and \citet{Brammer:2012aa}.  

We selected galaxies with a rest-frame {\it B} magnitude $ < 25$ and CLASS$\_$STAR $\leq 0.2$ in the redshift range stated above, based on their `zbest' redshift, which used the photometric redshift ($43\%$ of the sample) if no spectroscopic redshift ($57\%$) was available. We note that the use of photo-zs do not change our conclusions; we include the galaxies with only photo-zs to increase the statistics. There are 4479 galaxies cross-matched between the GOODS and 3D-{\it HST} catalogs.  Figure \ref{fig:zhist} shows the redshift distribution of the galaxy sample. \par

\subsection{Identification of AGNs}

Hard X-ray selection is one of the least biased and most efficient AGN selection techniques, as all but the most obscured systems can be detected, and relatively few non-AGNs are selected by this method. Additionally, X-ray AGNs have been shown to have no bias with respect to star formation rate \citep{Azadi:2016}. We select the AGN by cross-matching the positions of objects within $1^{\prime \prime}$ in the north and south Chandra Deep Field (CDF), using the CDFN 2 Ms point source catalog \citep{Alexander:2003aa,Xue:2016} and the CDFS 4 Ms catalog \citep{Xue:2011aa}. We use objects above an absorption-corrected, rest-frame 0.5-8 keV luminosity threshold $L_{\rm X}>10^{42}$ ergs s$^{-1}$ to reduce non-AGN contamination, and verify the redshifts match those of the galaxies. A total of 106 AGNs meet these criteria. \par

The color contamination of AGN host galaxies due to the central nucleus has previously been examined for these same HST-observed $z\sim 1$ galaxies \citep{Cardamone:2010aa}. By comparing the color before and after the point source was subtracted, \citeauthor{Cardamone:2010aa} concluded that the integrated galaxy color does not appear to be significantly affected by the central AGN light for these moderate luminosity AGNs, as long as the quasar light does not dominate its host galaxy. We have verified this applies to the AGN sample studied here by comparing the magnitudes of the point spread function (PSF) and galaxy from our morphological analysis. We therefore use the integrated color for all AGN host galaxies. Similarly, the bias the of stellar mass estimates for AGN hosts was found to be $<0.1$ dex \citep{Brammer:2009aa, Cardamone:2010aa}.\par

\begin{figure*}
\centering
\includegraphics[width=0.45\textwidth]{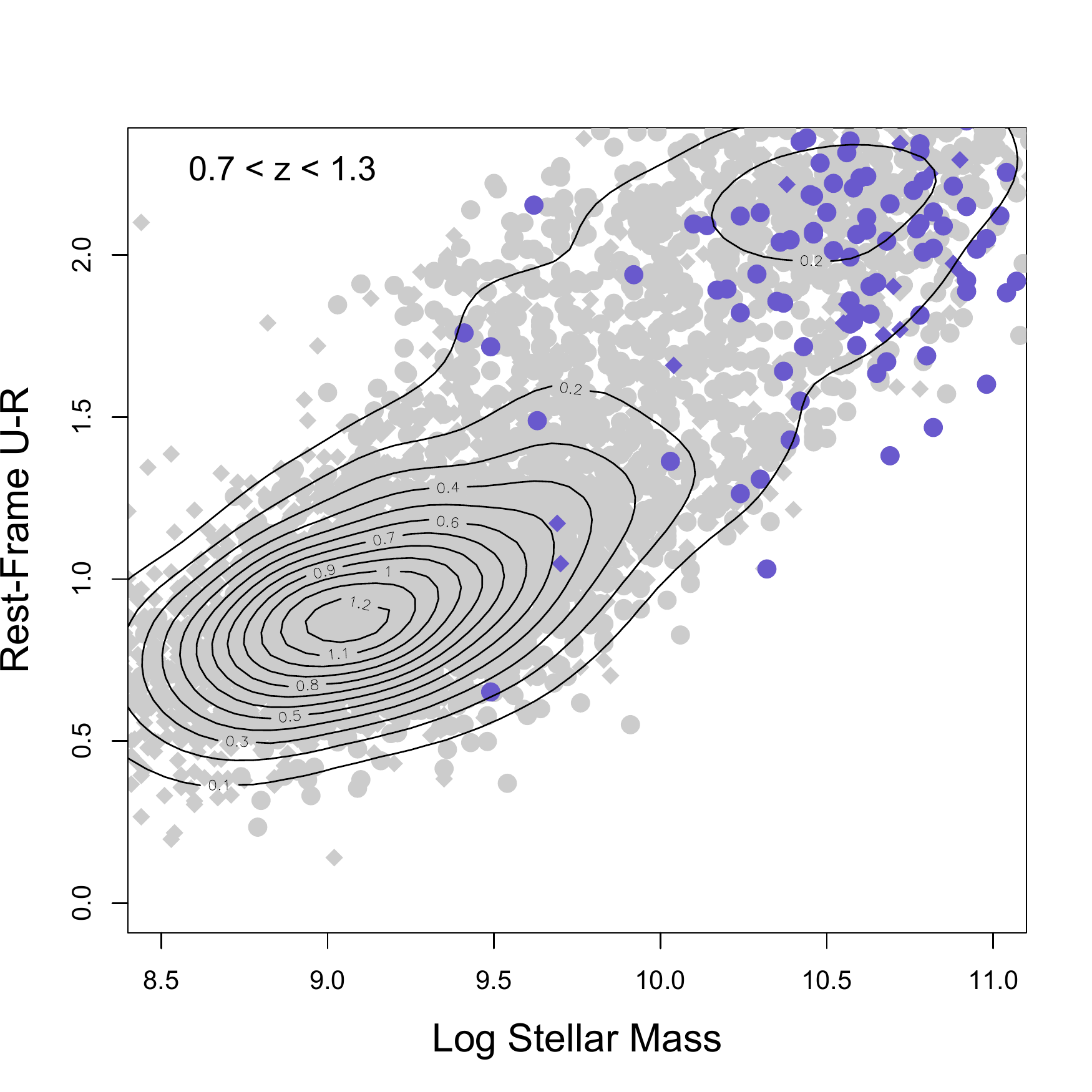}
\includegraphics[width=0.45\textwidth]{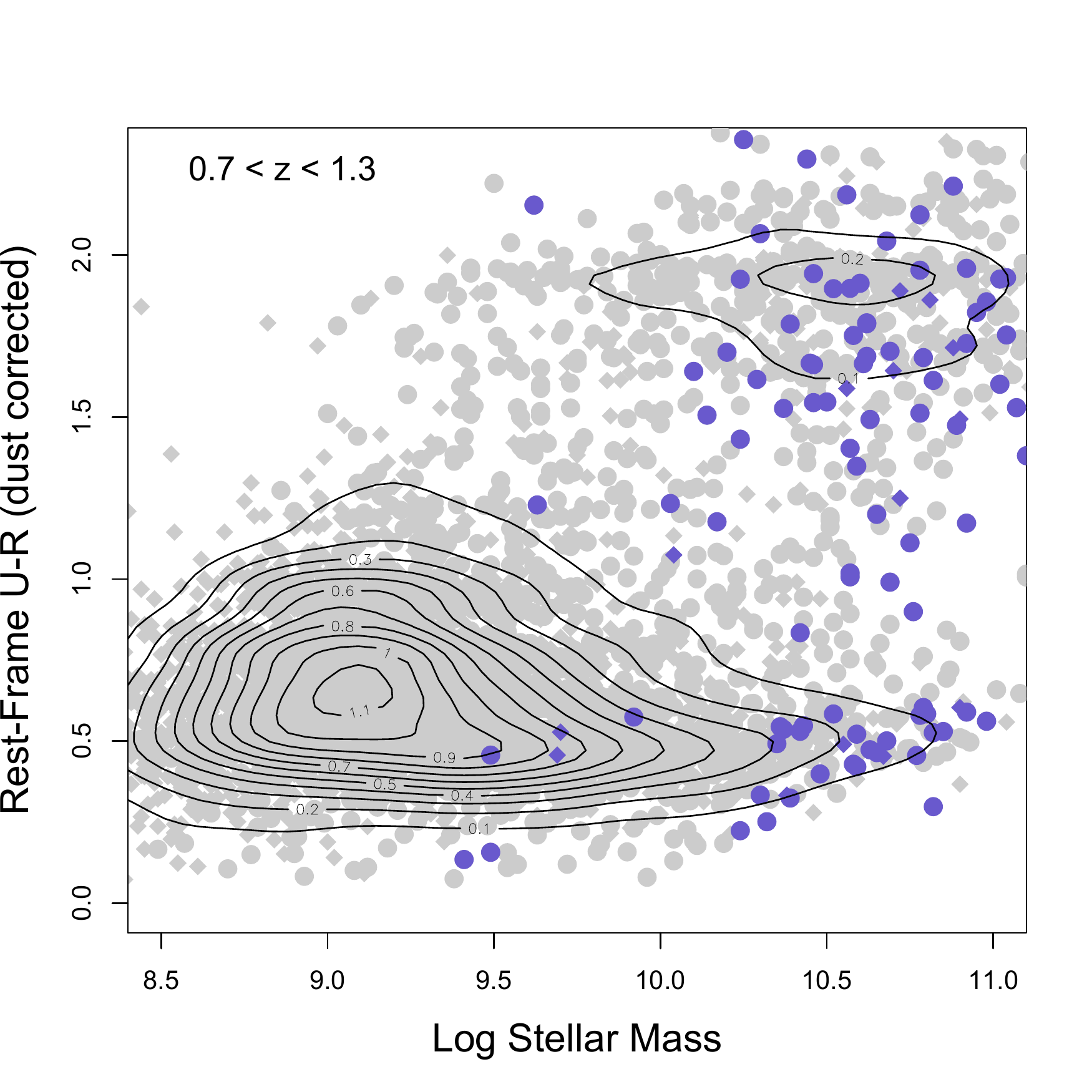}
\caption{Color-mass diagram of the full sample of 4479 galaxies and 106 AGN hosts within $0.7<z<1.3$, with the AGN hosts in purple. Galaxies with photo-zs only are shown with diamond points. The rest-frame U-R color is plotted on the left, and the dust-corrected color using the Calzetti et al. 2000 dust extinction law is plotted on the right, showing the well-known bimodal distribution of galaxies. The plots agree with similar studies at $z\sim 1$ \citep[e.g.,][]{Cardamone:2010aa, Brammer:2011aa}.}
\label{fig:uncorr}
\end{figure*}

\vspace{3mm}

Figure \ref{fig:uncorr} (left) shows the rest-frame U-R color versus log stellar mass of our 4479 galaxies. The gray dots and black density contours are associated with the inactive galaxy population, and the 106 AGN hosts are shown in purple.\par

It is known that obscuring dust can redden galaxies, so we correct the U-R colors for this reddening using the \citet{Calzetti:2000aa} extinction law

\begin{equation}
\Delta(U-R) = 0.65 A_{V},
\end{equation}

\noindent with $A_{V}$ extinction values from the 3D-{\it HST} catalog. After de-reddening, the red sequence and blue cloud are further separated, as seen in Figure \ref{fig:uncorr} (right). AGN host galaxy colors are distributed uniformly across the full range, while most galaxies are in the blue star-forming cloud, with a much smaller number in the red sequence and still fewer in the green valley. The red sequence is not as prominent as at $z\sim 0$ because there has been much less time since the peak of star formation for the red sequence to build up.\par

It appears that the AGNs preferentially lie in galaxies of large stellar mass. However several recent studies have characterized this as a selection effect, such that the probability that a galaxy hosts an AGN only depends on the underlying Eddington ratio distribution \citep{Aird:2010, Jones:2016, Azadi:2016}, and thus only a larger fraction of AGNs are found in massive galaxies. Additionally, we eliminate a larger fraction of the less massive AGN hosts by selecting AGNs with $L_{\rm X}>10^{42}$ ergs s$^{-1}$.

\hspace*{3mm} 

\section{Morphological Analysis}

\subsection{Analysis Method}

We performed 2D morphological analysis with GALFIT \citep{Peng:2002aa}, which uses $\chi^{2}$ minimization to fit parameters based on initial guesses for each component. An arbitrary number of components can be fit at once. We used PSFs from the 3D-{\it HST} survey, which were obtained by averaging between 35 and 200 carefully selected stars in each filter \citep{Brammer:2012aa}.\par

Initial parameter guesses, including integrated magnitude, half-light radius, axis ratio, and position angle, were obtained from the SExtractor catalogs \citep{Bertin:1996} produced by the GOODS survey \citep{Giavalisco:2004aa}. We made no assumption about \sersic\ index for each galaxy, using $n=2$ for the initial guess. If multiple components were fit, we input equal initial magnitudes for each component.\par

We first batch fit the inactive galaxies with a simple \sersic\ profile, then classified as disks those with \sersic\ index $n< 2$, and spheroids with $n \geq 2.5$. If $\chi^2_\nu > 2$ or convergence problems occurred, we then tried bulge/disk decomposition to determine the dominant component. If the fit was still poor ($\chi^2_\nu > 2$), the galaxies were individually fit with multiple components, with the brightest component determining the morphological type. The active galaxies were individually fit with a \sersic\ profile and PSF, or with a bulge, disk, and PSF. To simulate the effect of how unresolved features affect the fit, we fit a sample of inactive galaxies with an additional PSF, and found that the \sersic\ index typically changed by less than 0.1 (and no more than 0.5) from when fit with only a \sersic\ profile.

We additionally measured the asymmetry of each galaxy to estimate the fraction in our sample that are in ongoing mergers. Asymmetry was measured by calculating the fractional residuals of the image and the image rotated by $180^{\circ}$ \citep{Conselice:2000aa, Conselice:2003aa, Menanteau:2006, Shi:2009aa}:

\begin{equation}
A = \frac{\Sigma | I - I_{180^{\circ}}| - b}{2 \Sigma I},
\end{equation}

\noindent  where $b$ is the correction term due to the sky noise, calculated as $b=\sqrt{2}\sigma_{sky}N_{pix}$ ($N_{pix}$ is the area of the aperture; \citealt{Menanteau:2006}). We use an aperture radius of $3 R_{eff}$, where $R_{eff}$ is the best fit half-light radius found from GALFIT. Local galaxies with asymmetries $> 0.35$ almost always have disturbed morphologies from major mergers, however galaxies of the same morphology at higher redshift are more symmetric due to surface brightness dimming \citep{Conselice:2003aa, Shi:2009aa}. This can be corrected by decreasing the asymmetry threshold for mergers by $\delta A=0.05$ for $z\sim 1$ galaxies \citep{Conselice:2003aa}. Additionally, it was found that a significant fraction of galaxies above this asymmetry threshold can be contaminated with nearby stars or non-interacting galaxies \citep{DePropis:2007aa, Lopez:2009}. By visual inspection of all galaxies with $A>0.3$, this fraction is $\sim 50 \%$ for our sample. We therefore calculate the number of mergers as $50\%$ of the number of galaxies with asymmetries $>0.3$.

\subsection{Sample Selection}

\begin{figure*}
\includegraphics[width=0.8\textwidth]{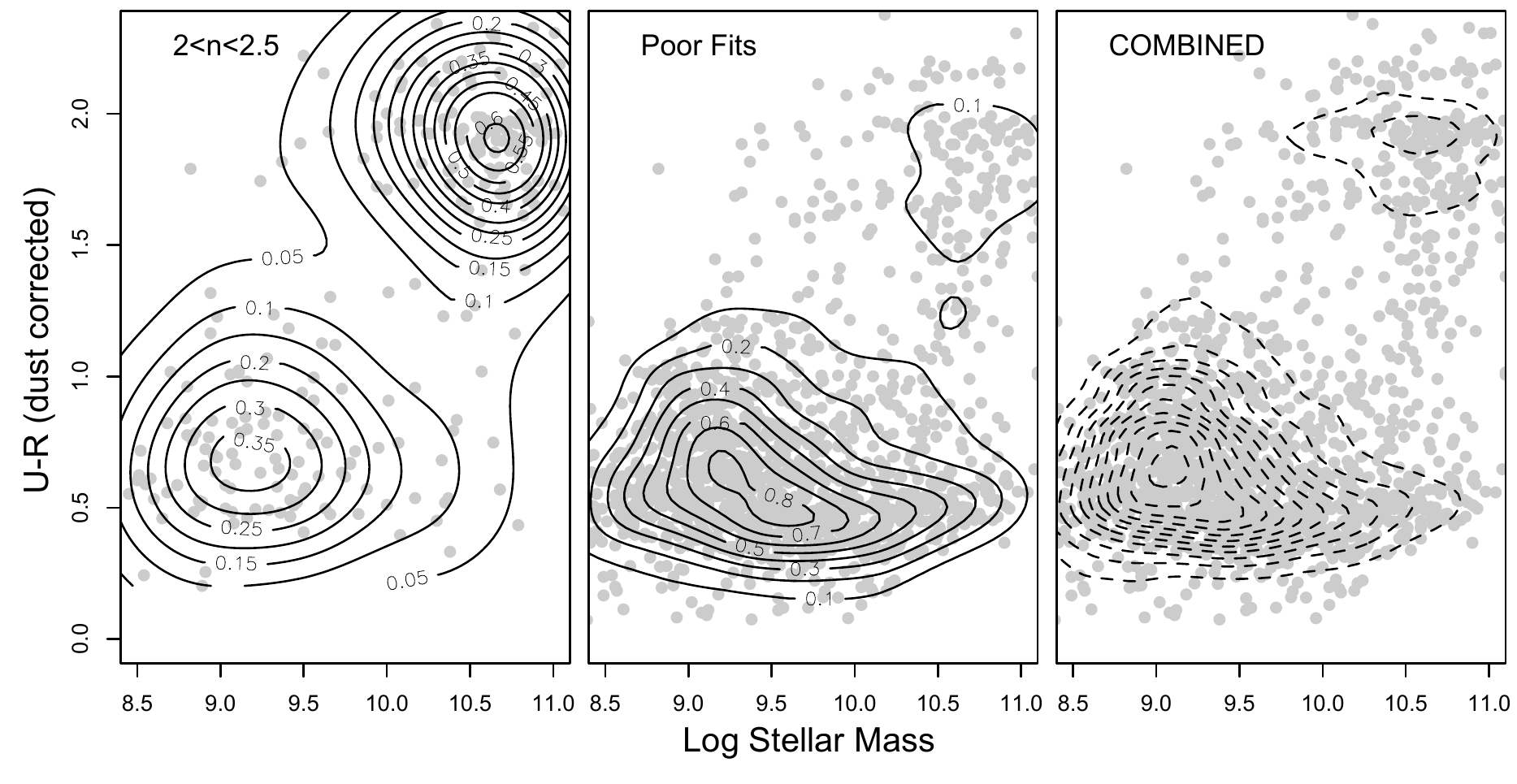}
\centering
\caption{Color-mass diagram of galaxies that were excluded from analysis (gray points), disaggregated into galaxies with intermediate \sersic\ index (left panel), which preferentially excludes a few bulge-dominated galaxies; galaxies with high reduced $\chi^{2}$ values for the fit, and/or overly large obtained effective radii (middle panel), whose color-mass distribution shows no bias; and the combined distribution of all cut galaxies (right panel) with the contours of the parent sample for comparison (dotted lines). Density contours normalized by the total number of points are overlayed in the left and middle plots (solid lines).}
\label{fig:cuts}
\end{figure*}

In this section we explain the cuts made to the galaxy sample to ensure the robustness of the morphological analysis. \par 

We eliminated fits with a $\chi^2_\nu>1.5$, and those with determined \sersic\ indices of $2 < n < 2.5$ to reduce contamination in each morphological category. If there are other sources in the field of view of the cutout, this could also cause error in GALFIT's index estimation, but this effect is reduced when removing fits with $\chi^2_\nu>1.5$. \par

For the AGN hosts that were fit individually, we excluded any with significant azimuthally symmetric residuals, indicating poor fits to the \sersic\ profile, or any irregular mergers. This cut out $\sim 20$ AGNs. \citet{Simmons:2008aa}, who analyzed 55,000 simulated {\it HST} ACS images to precisely define the limits within which galaxy morphologies can be well determined, found that fits to AGN host galaxies observed in GOODS are accurate as long as the host luminosity is greater than $1/4$ of the point source luminosity; based on this criterion, we excluded four more AGN host galaxies. Also, from following the results in \citet{Simmons:2008aa}, we make an effective radius cut $\sigma_{r_{e}}/r_{e} \geq 0.8$ for both inactive and active galaxy fits, which eliminates $\sim 88 \%$ of bad fits. 

For the measuring asymmetries, we follow \citet{Shi:2009aa} and disregard galaxies with $3 R_{eff}< 15$ pixels for a reliable measurement. This cuts out a negligible number of galaxies (40; $<2\%$ of each morphology type) from our merger fraction calculation.

Our final sample has 3010 galaxies and 83 AGN hosts, and we verified that it has the same color, stellar mass, and redshift distributions as the original sample. \par

Figure \ref{fig:cuts} shows the color-mass diagram of sources that were excluded from this analysis,
disaggregated into galaxies with intermediate \sersic\ indices and galaxies that are poorly fitted with a \sersic\ model. Galaxies with intermediate \sersic\ indices are evenly split between red sequence and blue cloud, as expected, which means this cut excludes a larger fraction of bulge-dominated galaxies. Galaxies with poor fits (i.e. $\chi^2_\nu>2$ or large effective radius) have similar colors and masses as the total galaxy sample, so their exclusion does not cause any bias in the color-mass distribution. \par

\subsection{Morphology Results}

\begin{figure*}
\centering
\includegraphics[width=.95\textwidth]{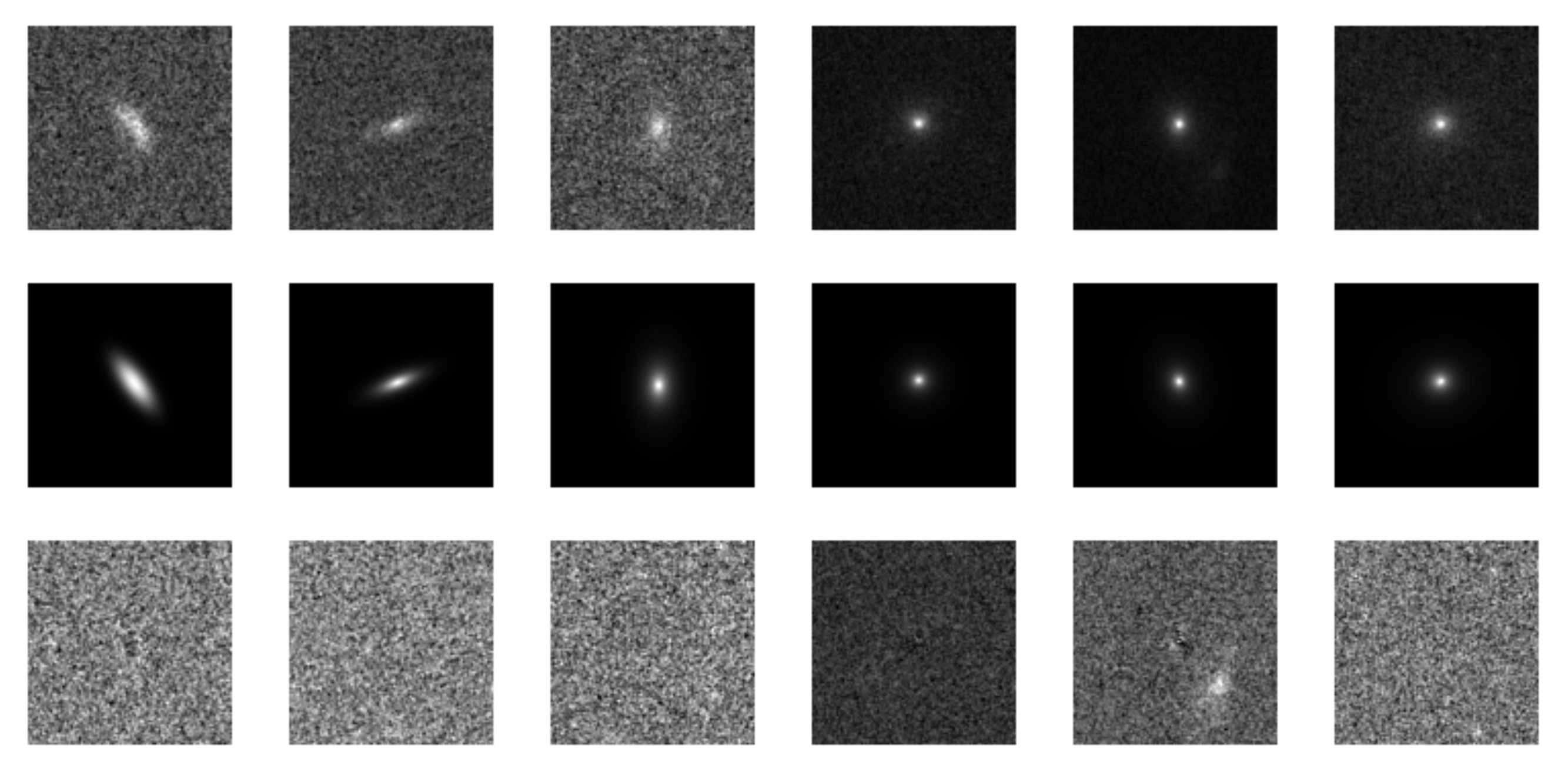}
\caption{Examples of 5 inactive galaxy fits and one AGN fit (right) from the GOODS fields, where each image is 3x3$^{\prime \prime}$. Top row shows the science images, the middle row shows the best-fit models generated by GALFIT, and at the bottom are the residuals. Fits were single \sersic\ profiles for the inactive galaxies, with \sersic\ indices of 0.5, 1.0, 1.5, 2.5, 3.3, respectively, and bulge/disk decomposition plus PSF for the AGN, with a dominant bulge component. }
\label{fig:fits}
\end{figure*}

\begin{figure}[h]
\centering
\includegraphics[width=.5\textwidth]{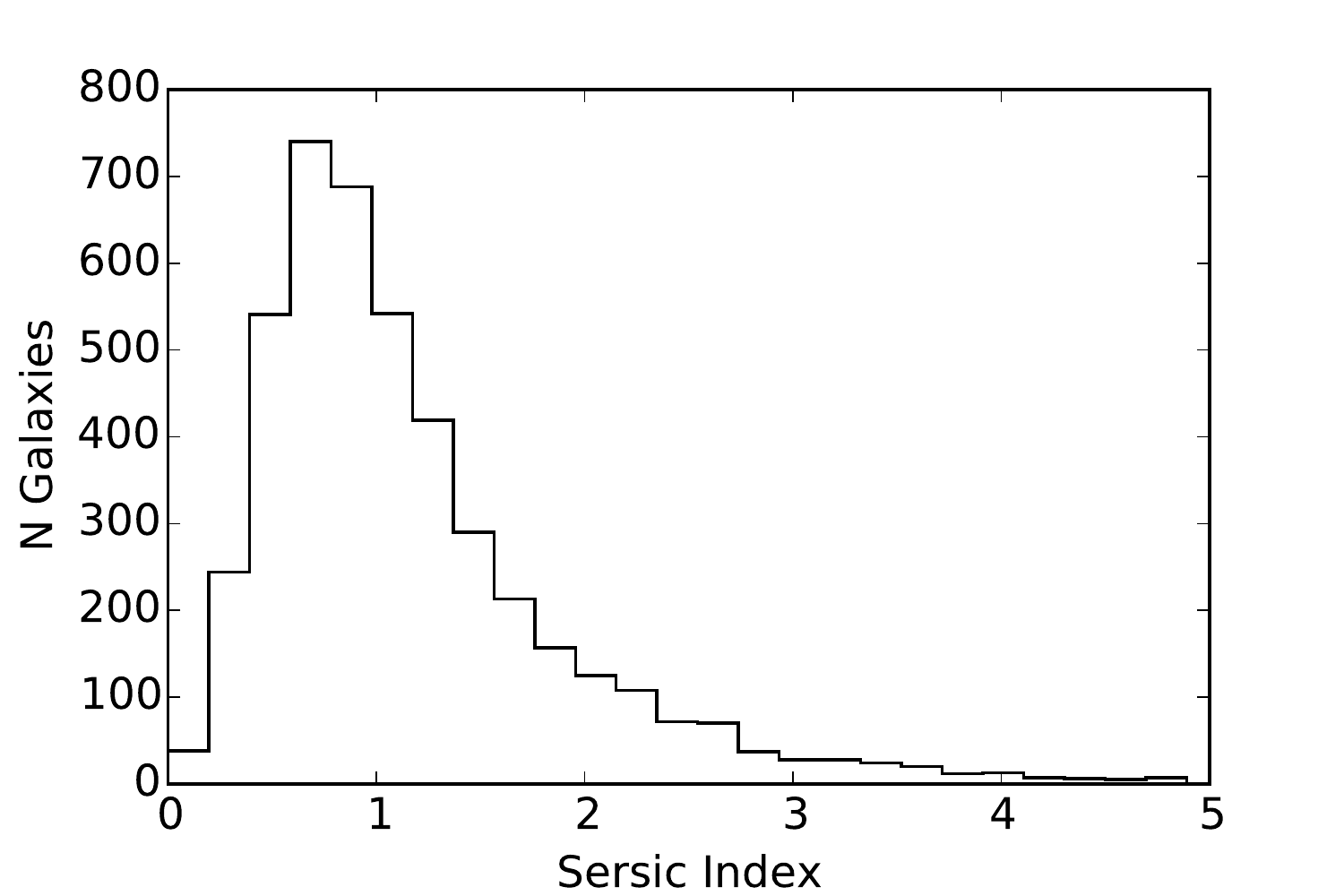}
\caption{Distribution of \sersic\ indices of all 3010 galaxies with acceptable fits. The majority of galaxies ($\sim 90 \%$) are disk-dominated (\sersic\ index $<$ 2).}
\label{fig:ndist}
\end{figure}

\begin{figure}
\centering
\includegraphics[width=.5\textwidth]{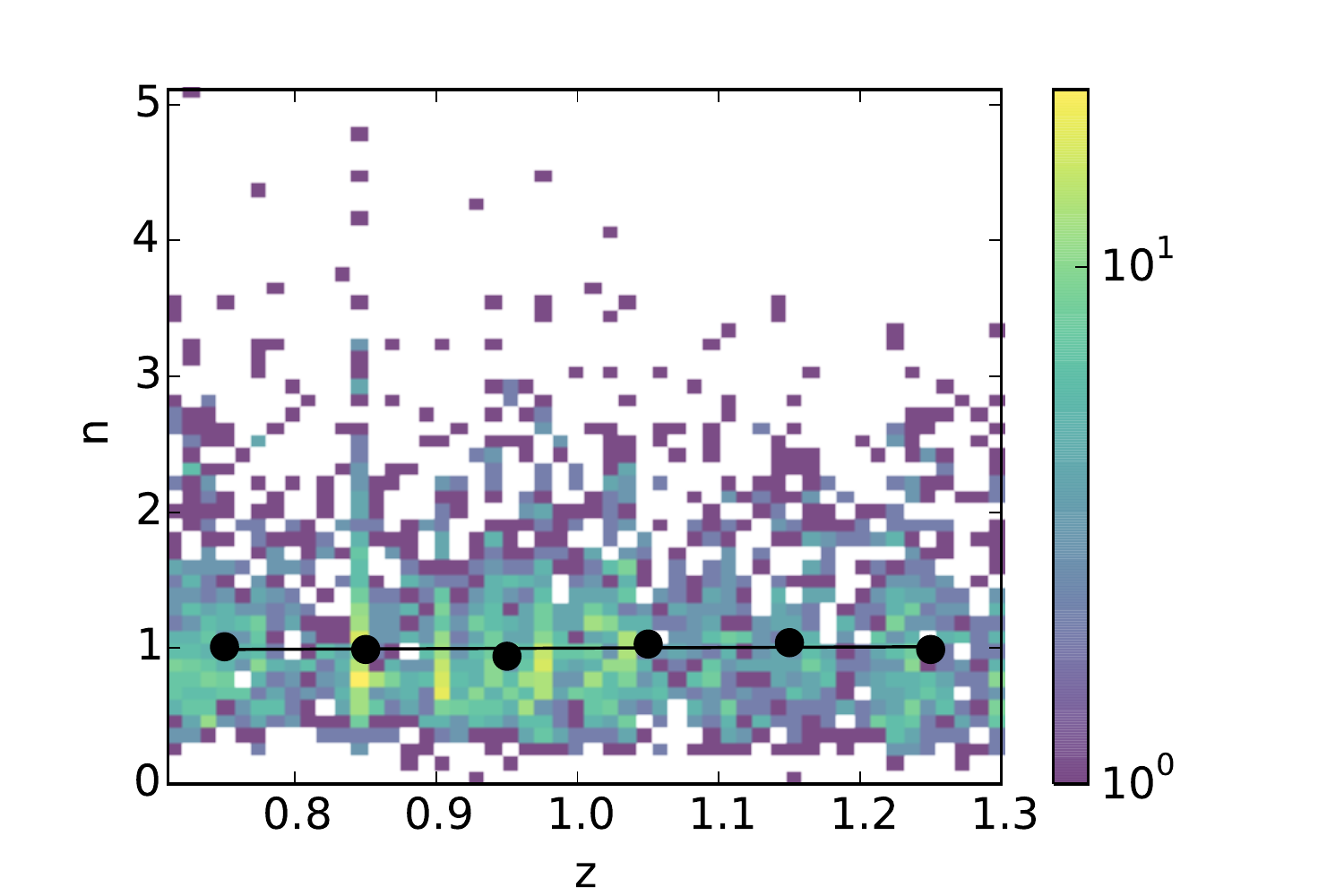}
\caption{2D histogram of morphological fits of all 3010 galaxies as a function of Sersic index, $n$, and redshift, $z$. The number per bin are shown by the color bar. We verify there is no systematic dependence of morphology fit on redshift. Median \sersic\ index values per redshift bin are shown in black points, and the black line shows the least square line fit, which has a slope of $0.04$.}
\label{fig:nvsz}
\end{figure}

Figure \ref{fig:fits} shows the image data, model, and residuals for example galaxies of the different morphological types, and the \sersic\ index distribution is shown in Figure \ref{fig:ndist}. We find that a large majority of the galaxies ($\sim 90\%$) at $z\sim 1$ are disk dominated, and there is no redshift dependence of morphology, shown in Figure \ref{fig:nvsz}. These results are generally consistent with other morphological analyses of galaxies in GOODS \citep{Bundy05,vanderWel12}; we compare our results to those in more detail in the Appendix \ref{app:comp}.

\section{Color-Mass and Morphology}

\begin{figure*}
\centering
\includegraphics[width=.44\textwidth]{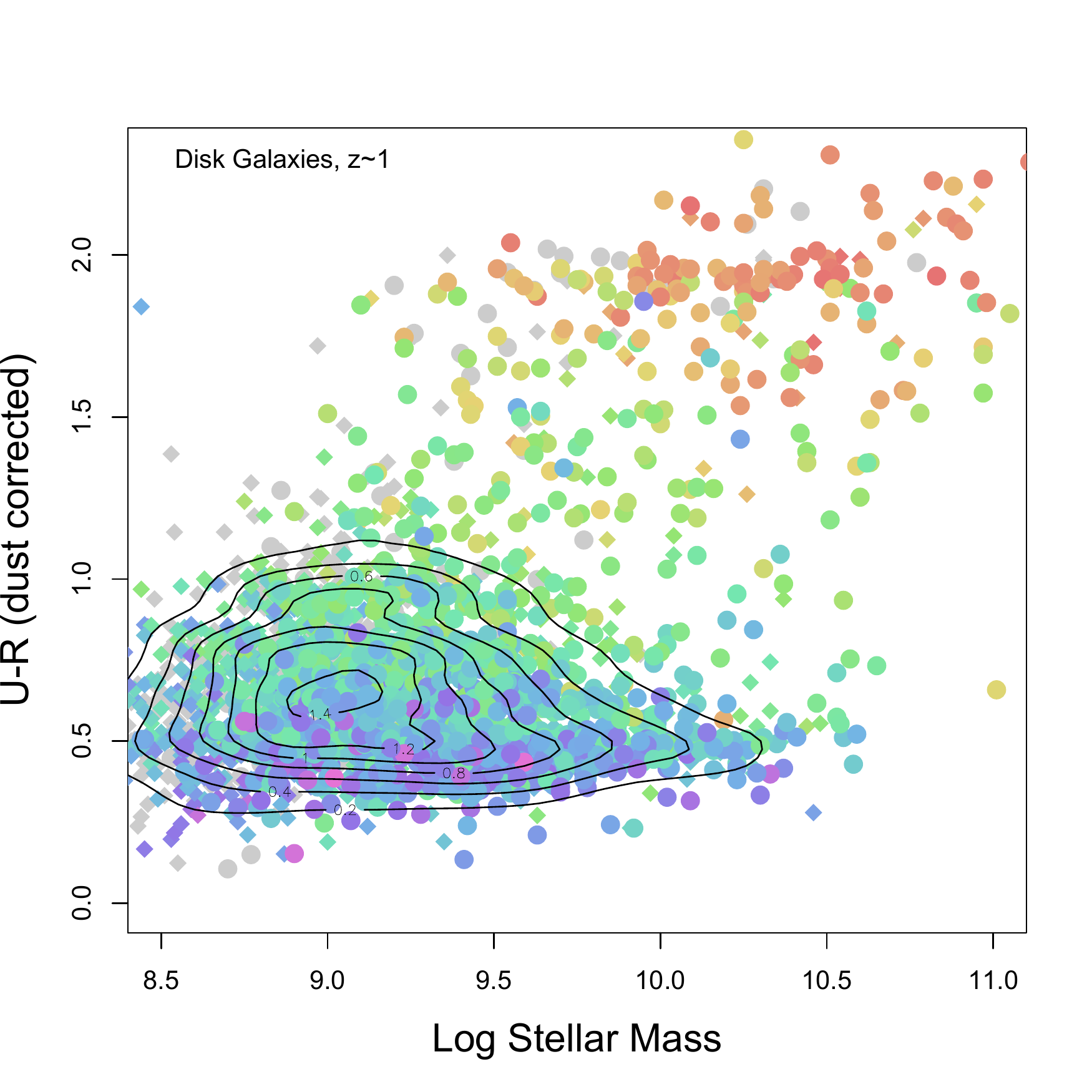}
\includegraphics[width=.44\textwidth]{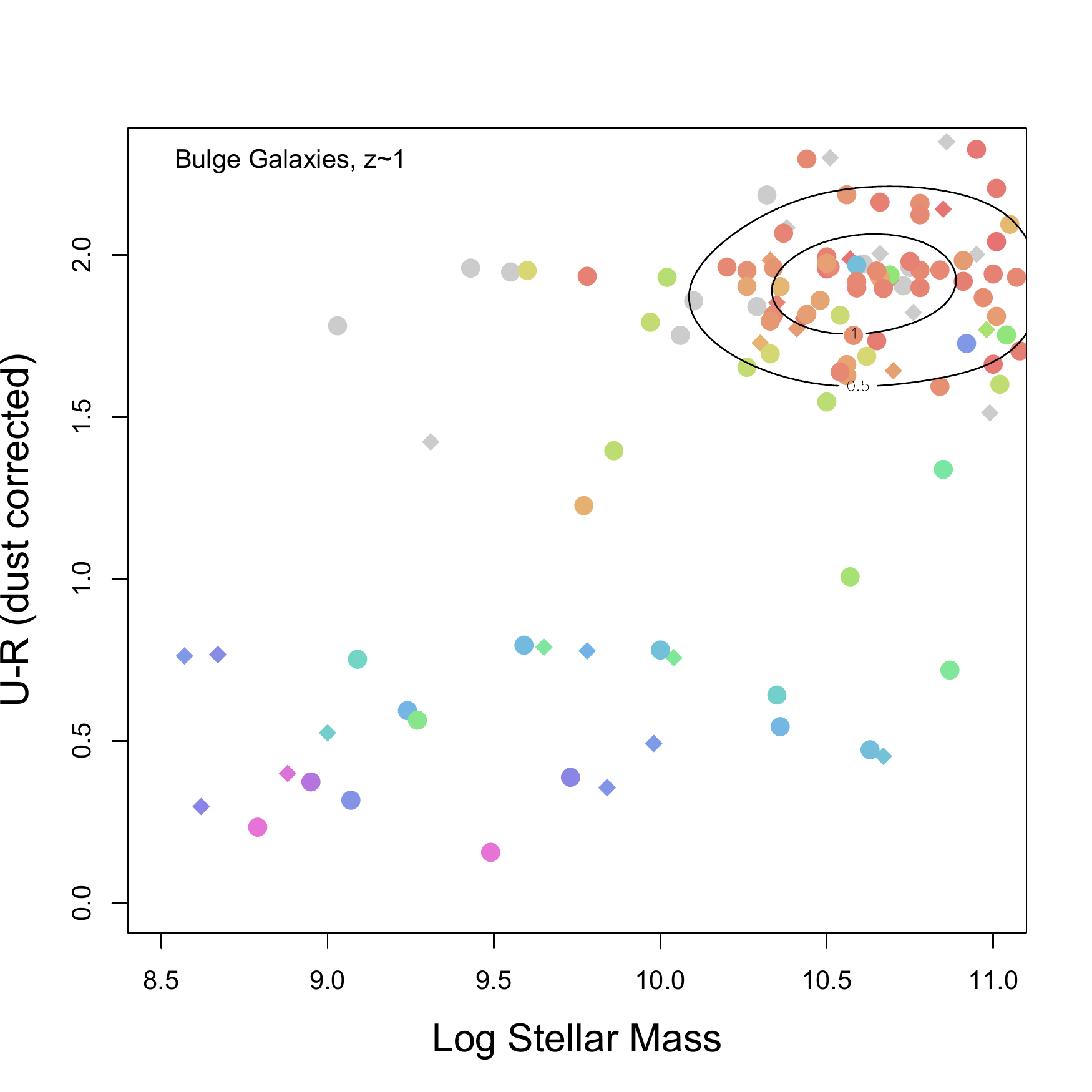}
\includegraphics[width=.08\textwidth]{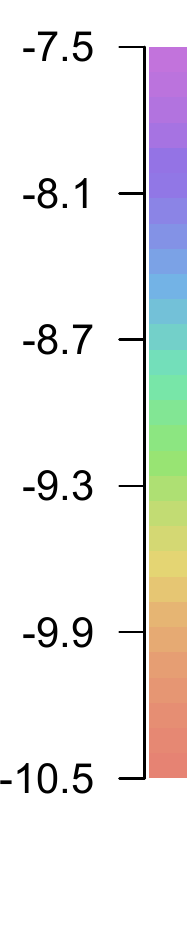}
\centering
\caption{Dust corrected U-R color vs. log stellar mass for $z\sim 1$ galaxies, shown separately for disk-dominated morphology (left; \sersic\ index n$<$2), and bulge-dominated morphology (right; n$>$2.5). The color of each point corresponds to the galaxy's specific star formation rate from \citet{Whitaker:2014aa} (shown by the colorbar at right labelled by the corresponding log of the specific star formation rate), and the shape corresponds to the redshift type (circle for spectroscopic and diamond for photometric). The unimodal distribution of disk galaxies extending toward higher masses suggests a slow quenching process, while the low number of spheroid galaxies in the green valley suggests a faster quenching time.}
\label{fig:morph}
\end{figure*}

Figure \ref{fig:morph} shows color-mass plots for disk and bulge-dominated inactive galaxies at $z \sim 1$ with the associated density contours. The colors of the points correspond to the specific star formation rates of the galaxies from 3D-HST \citep{Whitaker:2014aa}, showing the decline in star formation as galaxies move to the red sequence. The disk-dominated galaxies peak in the blue cloud while the bulge-dominated galaxies peak in the red sequence, unsurprisingly, but both span the entire range in color. The demographics of galaxies in the blue cloud, green valley, and red sequence for each morphology are summarized in Table \ref{table:dem}. We define green valley galaxies as galaxies with dust-corrected rest-frame U-R colors between 1 and 1.5 mag; most are transitioning to the red sequence but at least 15$\%$ (estimated from visual inspection of a random sample of 100 galaxies) are simply blue disk galaxies with a red bulge. \par

\begin{table}[h]
\begin{tabular}{ p{4cm}p{1cm}p{2cm}  }
 \multicolumn{3}{}{} \\
 Galaxy Sample & $N$ & $\%$ Population \\
 \hline
\hline
 Early-type, blue cloud      &  26       & 20.6$\%$ \\
 Early-type, green valley   &   5        &   4.0$\%$ \\
 Early-type, red sequence & 95       & 75.4 $\%$\\
\hline
 \emph{Early-type, all}       &126 &100.0$\%$ \\
     & & \\
\hline
\hline
 Late-type, blue cloud        &   2257  &  85.1$\%$\\
 Late-type, green valley     &  193    &  7.3 $\%$  \\
 Late-type, red sequence   &  201    & 7.6 $\%$\\
\hline
 \emph{Late-type, all}         & 2651  &100.0$\%$ \\
\end{tabular}
\caption{Demographics of inactive galaxies by color and morphology.}
\label{table:dem}
\end{table}
\vspace{3mm}

The left-hand plot of Figure \ref{fig:morph} shows that the number of disk-dominated galaxies peaks in the blue cloud but extends up to the red sequence, similar to what \citet{Schawinski:2014aa} found at $z\sim 0$, namely, that the disk galaxies experience a gradual decline in star formation. The disks span the color-mass diagram, but there is no bimodality to suggest a rapid transition through the green valley. This is in agreement with other recent studies of the star formation histories of low redshift disk galaxies \citep{Wolf:2009, Masters:2010, Tojeiro:2013aa, Lopes:2016aa}.\par

In contrast, bulge-dominated galaxies (Fig \ref{fig:morph}, right) peak in the red sequence with very few in the green valley. This closely resembles what is seen at $z\sim 0$ \citep{Schawinski:2014aa}, in which spheroidal galaxies quench more rapidly. We find that only $\sim 2 \%$ of green valley (inactive) galaxies are bulge-dominated. However, the fraction of spheroids that are blue is much higher than what was found locally; $\sim 21\%$ of early-type galaxies are in the blue cloud, compared with $\sim 5\%$ at $z\sim 0$ \citep{Schawinski:2014aa}. It should be noted that in Schawinski et al. the galaxy morphologies were classified by eye via the Galaxy Zoo citizen science project \citep{Lintott:2008aa}, and that intermediate galaxy shapes were excluded from the analysis, similar to our exclusion of intermediate \sersic\ indicies. These excluded galaxies were mainly disks with bulges, so we are comparing similar sub-samples. 

There are several possibilities for the origin of the blue spheroidal population besides forming from major mergers. One possibility is from violent disk disruption, which would compactify disks into small star-forming ellipticals \citep{Dekel:2013aa}. However, these compact `nugget' galaxies are generally observed at higher redshift, $z=2-3$, and the their effective radii are of order 1 kpc \citep{Dekel:2013aa}. The blue ellipticals we measured here have effective radii of several kpc, of the same order as elliptical galaxies at $z \sim 0$. Additionally, some spheroidal galaxies could trickle down from the red sequence after a spontaneous burst of star formation, perhaps after a minor merger. Recent studies investigating this population have not converged on a single evolutionary scenario for the blue bulge population, but have found evidence that there are contributions from both major mergers as well as `rejuvination' \citep{Smethurst:2015aa, Tojeiro:2013aa, Lopes:2016aa}.  

\begin{figure}
\centering
\includegraphics[width=.45\textwidth]{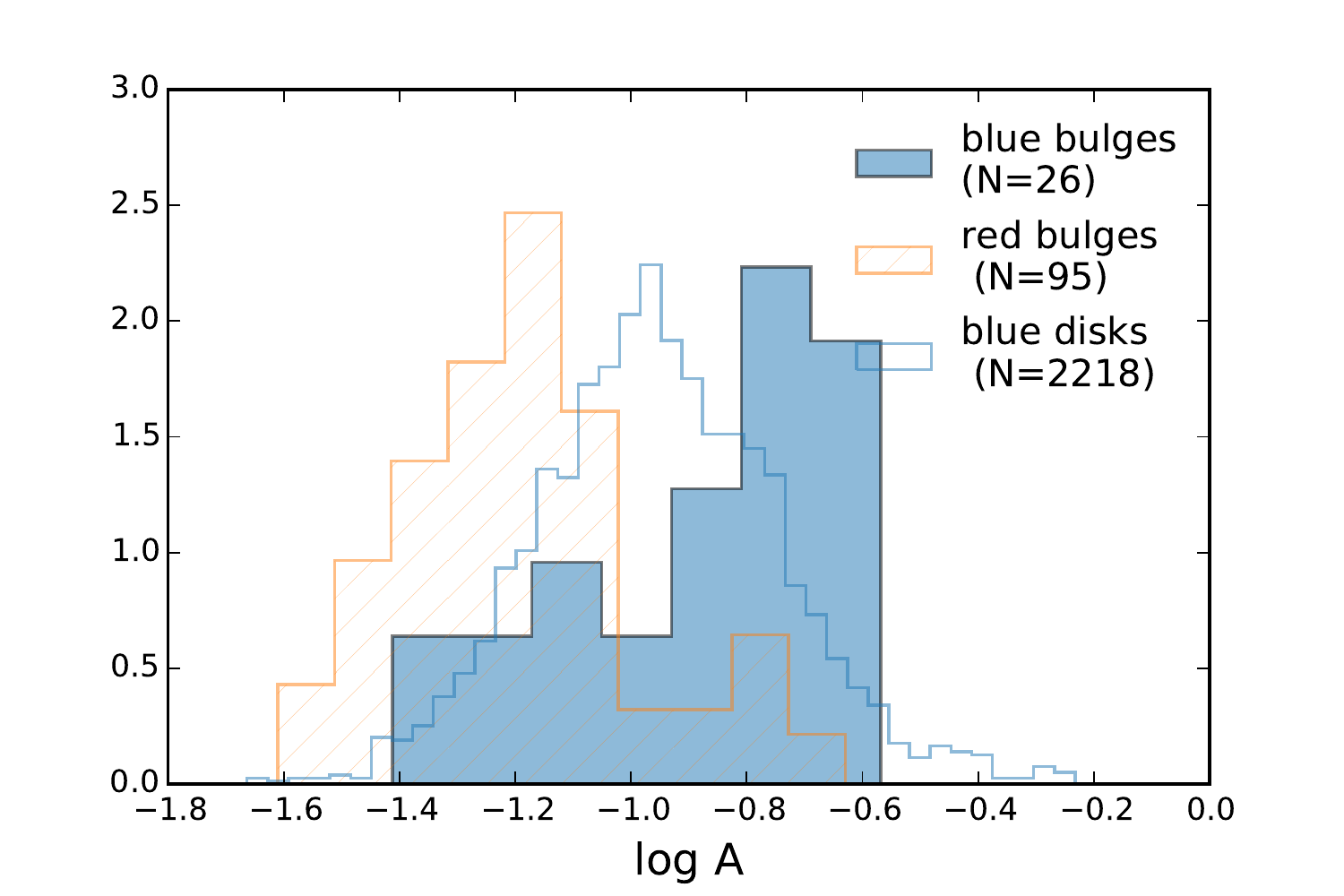}
\caption{Distributions of the asymmetries of each morphological type (normalized such that the integral over the range is 1), showing that the blue bulge population is more asymmetrical than the disks or red spheroids. Median values for the blue bulges, blue disks, and red bulges are 0.16, 0.11, and 0.06, respectively.}
\label{fig:asymm}
\end{figure}

We investigated further by measuring the asymmetry of the blue spheroids. If they formed from recent major mergers they should have a higher fraction of disturbed morphologies compared with the blue cloud disks and red sequence ellipticals. We found that is indeed the case, shown by Figure \ref{fig:asymm}. The offset of asymmetry between the blue elliptical peak and the red elliptical peak is evidence that there is at least a significant fraction of blue ellipticals that are associated with mergers, agreeing with recent findings \citep{Smethurst:2015aa, Lopes:2016aa}. By visual inspection, it was clear that the galaxies with $A>0.15$ were significantly asymmetric.

The merger fraction, along with the relative timescales between the change in morphology from a major merger and the change of color, should dictate the fraction of merger-formed blue ellipticals that we see. Mergers were more common at $z\sim 1$ than at $z\sim 0$, with an observed dependence that goes roughly as $\propto (1+z)^{3\pm 1}$ \citep{Lotz:2008ab}, and so the increase in the fraction of blue spheroids by a factor $\sim$4 or 5 that we see is consistent within error of what would be expected if they were formed by major mergers. By measuring the asymmetries of all blue cloud galaxies, we calculated a merger fraction for our sample of $\sim 0.016$. This is low compared to other merger fraction calculations in the literature; however, different galaxy selections, aperture areas, and contamination correction factors make direct comparisons difficult \citep{Lotz:2011}. We emphasize that our calculation is an order of magnitude estimate. 

The fraction of blue ellipticals that we find in the blue cloud and green valley is ($0.014\pm0.003\%$), within error of the calculated merger fraction. This would be the fraction expected if these galaxies are the result of major mergers. The timescale for a galaxy's morphology to relax is estimated to be a few hundred Myrs \citep[e.g.,][]{Conselice:2006b,Lotz:2008ab} while the quenching timescale for early-type galaxies is of the same order or longer, $\leq 1$ Gyr \citep{Schawinski:2014aa, Smethurst:2015aa}, making it difficult to predict the number of relaxed blue spheroids for a given merger rate. However our findings are consistent with the picture in which the morphology changes before the onset of rapid quenching for a significant fraction of galaxies that have undergone a major merger. 

In summary, given our calculated merger fraction and observed number of blue ellipticals and their asymmetries, we argue that they are mostly late-stage mergers. Assuming this, and the fact that the estimated timescales of morphological relaxation is not significantly shorter than the timescale for quenching, our results are consistent with the scenario where quenching is delayed after galaxies undergo a major merger, possibly due to delayed black hole accretion. However this depends on exact merger rates and the relative timescales between a galaxy's morphology change and color change, and we cannot currently break these degeneracies.

\begin{figure}
\centering
\includegraphics[width=.45\textwidth]{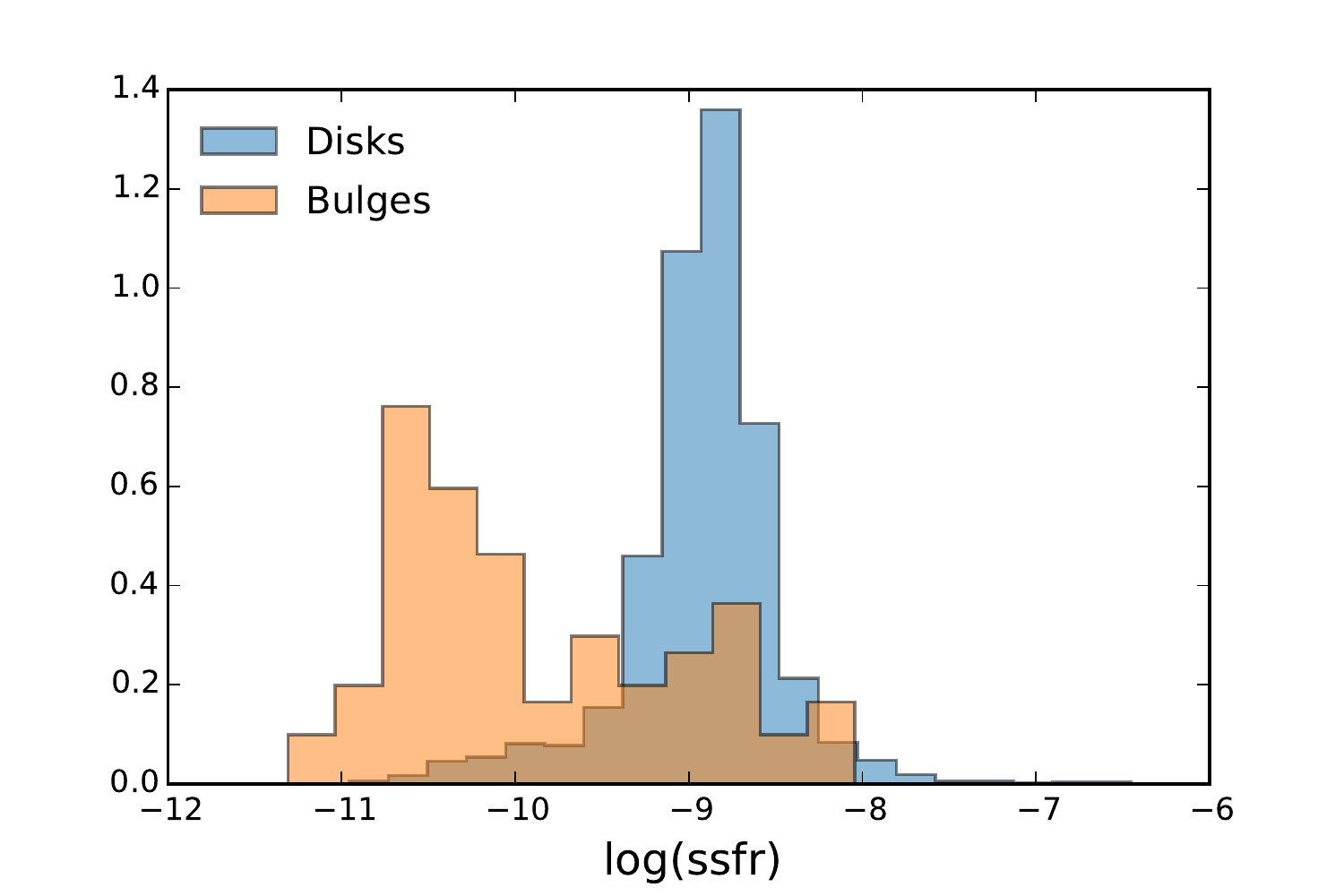}
\caption{Distributions of the log of the specific star formation rates for bulge galaxies ({\it orange}) and disks ({\it blue}), normalized such that the integral over the range is 1. The bulges have more of a bimodal distribution and primarily peak at a lower specific star formation rate than the disk-dominated galaxies. The secondary peak in the star forming region for bulge galaxies may suggest that quenching is delayed for some time after galaxies undergo a major merger.}
\label{fig:lssfr}
\end{figure}

Figure \ref{fig:lssfr} shows the normalized distributions of specific star formation rate estimates for both morphology types. The disk galaxies peak at a higher rate than bulges, confirming the association of disks with long-term star formation and of bulges with recent quenching. The bulge-dominated galaxies have more of a bimodal distribution as can also be seen on the color-mass diagram. The peak at higher specific star formation rate again suggests blue bulge-dominated galaxies stay blue for some time following a merger, implying that quenching is delayed after galaxies undergo a major merger. This would be expected if AGNs were responsible for quenching, as it takes time for gas to funnel to the center of the galaxy before igniting the AGN.

\subsection{AGN Host Galaxies}

If AGNs are responsible for the quenching we should expect some AGN hosts to be in the blue cloud, because it takes several hundred Myr for a stellar population to age to green valley colors \citep{Schawinski:2009ab}. Additionally, if the initial AGN luminosity is the highest, as some models claim \citep[e.g.,][]{Hopkins:2007aa}, we should see AGNs with higher Eddington ratios closer to the blue cloud than to the red sequence.\par

\begin{figure*}
\centering
\includegraphics[width=.45\textwidth]{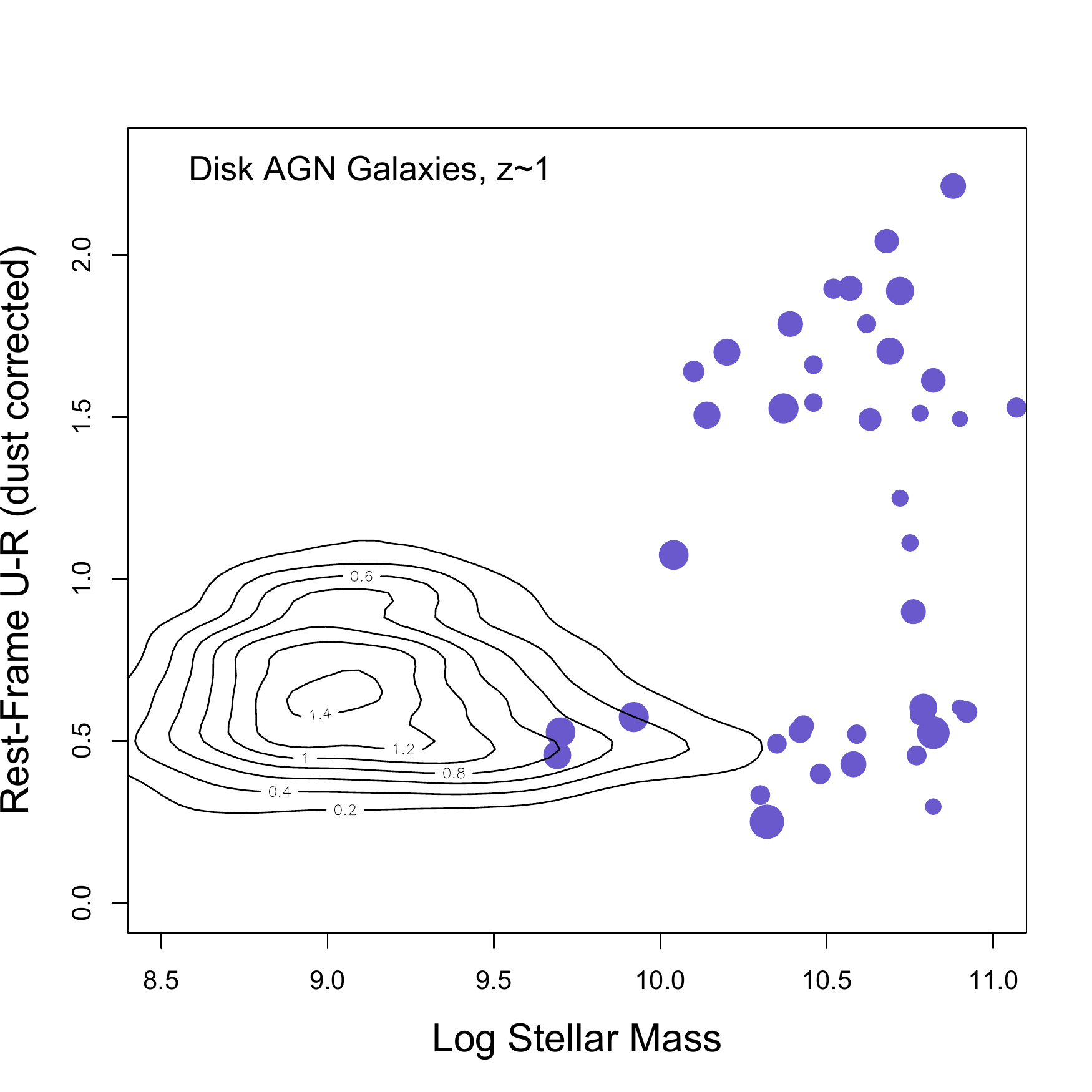}
\includegraphics[width=.45\textwidth]{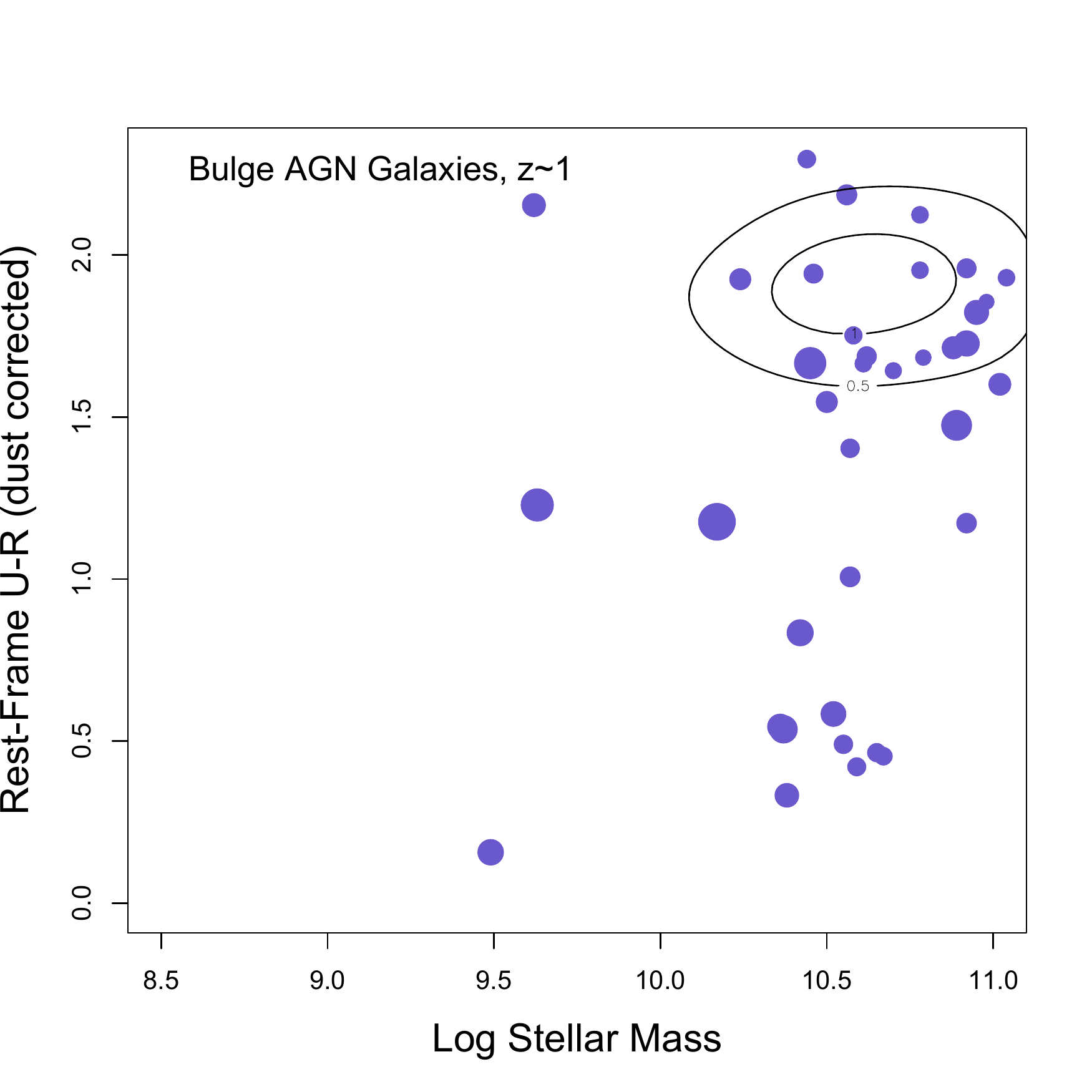}
\caption{Color-mass diagram for the host galaxies of AGNs, separated into disk-dominated galaxies (left) and bulge-dominated galaxies (right). The contours represent the (much larger) distribution of inactive galaxies, and the size of the points roughly correspond to the black hole Eddington ratios. The distribution of AGN hosts is roughly uniform in color space for both morphologies. They are also approximately equal in number, showing that mergers cannot be the only or main trigger of AGNs.}
\label{fig:agnmorph}
\end{figure*}

\begin{figure*}
\centering
\includegraphics[width=.8\textwidth]{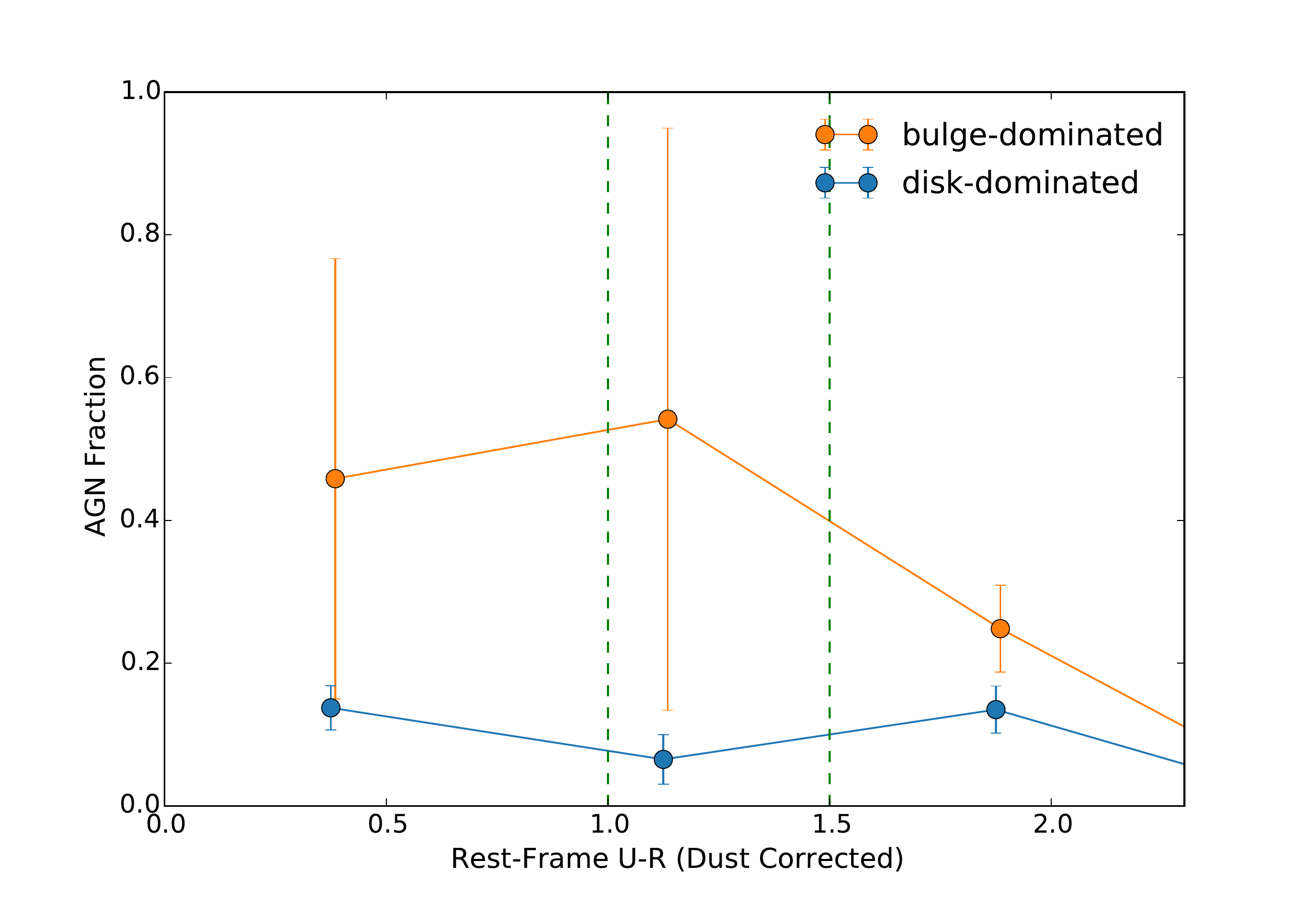}
\caption{The fraction of galaxies above $M_{*}=10^{10} M_{\odot}$ that host AGNs versus rest-frame U-R color for both bulge-dominated (orange) and disk dominated galaxies (blue). The green valley lies in between the two vertical dotted lines, and the normalization is corrected for the number of excluded galaxies and AGN. Although not statistically very significant, the fraction of bulge-dominated AGN hosts is larger than for disk galaxies in the blue cloud and green valley. This suggests that AGN feedback is not significant for galaxies that have not undergone a major merger, but may play a role in quenching for galaxies that have. }
\label{fig:frac}
\end{figure*}

Separating the the AGN hosts into disk-dominated and bulge-dominated, their color-mass diagrams are shown in Figure \ref{fig:agnmorph} along with the density contours of the inactive galaxies for comparison.  The size of the points corresponds to 0.5-8 keV X-ray luminosity divided by stellar mass, which should approximate the Eddington ratio if the bolometric correction and the ratio of black hole to stellar mass are roughly constant (the bolometric correction was empirically found for most obscured GOODS AGNs to be $\sim 40$ in \citealt{Simmons:2011aa}). We also plot the fraction of galaxies of each morphological type hosting an AGN above a stellar mass threshold of $10^{10} M_{\odot}$ for various color bins in Figure \ref{fig:frac}. We chose the mass threshold so that AGN incompleteness at low stellar mass does not affect the conclusions. As there can be selection issues with AGN fractions, we looked at how restricting our sample to high $L_{X}$ or $L_{X}/M_{*}$ would affect Figure \ref{fig:frac}; as there was effectively no qualitative change except larger error bars, we include the full sample of AGNs in Figure \ref{fig:frac}. The morphological demographics for the AGN host galaxies are given in Table \ref{table:agndem}.\par

The absolute distributions of disk-dominated and bulge-dominated AGN hosts are similar, i.e, that both uniformly span the color space at higher masses, and they are also approximately equal in number. This suggests both that mergers may trigger some AGNs but cannot trigger more than about half the AGNs at $z\sim 1$, since half are disk dominated indicating no recent major merger. Although we are limited by small statistics, there is no evidence for a difference of stellar mass between the two populations that was found at $z\sim 0$ \citep{Schawinski:2010}. A Kolmogorov-Smirnov test confirms their distributions in the color-mass plane are not significantly different (\emph{p}-value 0.08 for the null hypothesis). On the face of it, this would suggest that AGNs have little to do with morphology, and by extension, with mergers. Additionally, there is no systemic discrepancy of the Eddington ratios between the two morphologies (with a \emph{p}-value of being drawn from different distributions $\sim$0.65).\par

Considering the fraction of galaxies that host AGNs, there are marked differences with galaxy morphology. The fraction of disk galaxies hosting an AGN is small and shows little color dependence, as seen in Figure \ref{fig:frac}. This implies that secular triggering of significant black hole growth is relatively rare. The trigger for these disk-dominated AGNs must not be a major merger, as their disk morphologies last to the red sequence. Also, because we find no excess of these hosts in the blue cloud or green valley we conclude that AGN feedback cannot be the major driver of galaxy evolution for disk galaxies. 

For the minority of galaxies that undergo a major merger and become bulge-dominated, we see a higher incidence of AGN activity while the host galaxy colors are still blue, although the significance is limited by the small sample size. This does imply a possible role for feedback in moving them from the blue cloud into the green valley, as was found by \citet{Schawinski:2014aa}. However, there is no evidence for a trend in the black hole's Eddington ratio with respect to the color. While this may challenge the idea that luminous AGNs quench star formation, our AGN sample excludes the rare high-luminosity quasars because of the small survey volume. For our sample of moderate-luminosity AGNs, we find no significant evidence that AGN hosts evolve differently from inactive galaxies, agreeing with other recent studies on AGN host galaxies \citep{Goulding:2014}.\par

Lastly, we note that if the AGN phase is continuous it must last long enough to reach the red sequence, namely $\sim 1$ Gyr, as there are AGNs of both morphological types with red host galaxies. However, recent studies have suggested AGNs flicker on and off within the time scale of traversing the green valley \citep[e.g.,][]{Cavaliere:1989, Schawinski:2015aa}, or even much shorter time scales \citep{Hickox:2013aa}. In that case, the location of an AGN in the color-mass diagram is essentially random, i.e., unrelated to the age of the stellar populations. This is what we see, as the AGN colors are distributed across the full color space. Of course, the density of AGNs relative to galaxies does change across the diagram and so the flickering cannot be random across the entire galaxy lifetime. Rather, the AGN flickering should occur after quenching starts, as the stellar population is aging. This should become more clear with larger samples.

\begin{table}
\begin{tabular}{ p{4cm}p{1cm}p{2cm}  }
 \multicolumn{3}{}{} \\
 AGN Sample & $N$ & $\%$ Population \\
 \hline
\hline
 Early-type, blue cloud      &  10       & 25.0$\%$ \\
 Early-type, green valley   &   7       &   17.5$\%$ \\
 Early-type, red sequence & 23       & 57.5 $\%$\\
\hline
 \emph{Early-type, all}       &40 &100.0$\%$ \\
     & & \\
\hline
\hline
 Late-type, blue cloud        &   19 &  46.3 $\%$\\
 Late-type, green valley     &  5  & 12.2 $\%$  \\
 Late-type, red sequence   &  17    & 41.5 $\%$\\
\hline
 \emph{Late-type, all}         &41  &100.0$\%$ \\
\end{tabular}
\caption{Demographics of AGN host galaxies in the blue cloud, green valley, and red
sequence for each morphology type. }
\label{table:agndem}
\end{table}
\vspace{3mm}

\section{Summary}

We have measured morphologies of 3000 galaxies in the CANDELS/GOODS north and south fields to examine the dependence of major mergers on the evolution of $z\sim 1$ redshift galaxies. Taking advantage of deep, multi-wavelength data available in these fields, we have plotted the distributions on the color-mass diagram for both disk-dominated and bulge-dominated galaxies. We then compared to the X-ray selected AGN hosts in the same field to study the role of mergers in triggering AGNs, and the role of AGN feedback in quenching star formation.

Our main conclusions are as follows:

\begin{itemize}
\item{
	The distribution of disk-dominated galaxies on the color-mass diagram is unimodal, but spans the entire diagram up to the red sequence. This suggests that these galaxies evolve gradually, slowly exhausting gas via secular processes.
}

\item{
	The lack of bulge-dominated galaxies in the green valley suggests faster quenching for galaxies that have undergone a major merger. 
}

\item{
Disk-dominated AGN hosts range from the blue cloud to the red sequence, but there is little evidence for AGN feedback. The trigger for these AGNs must be a mechanism that does not disrupt the disk.
}

\item{
There is a larger fraction of bulge-dominated AGN hosts in the blue cloud and green valley, allowing for the possibility of AGN feedback occasionally playing a role in quenching.
}

\item{
	There are more blue ellipticals at $z\sim 1$ than at $z\sim 0$. We infer that this is due at least partially to the fact that mergers were more common at higher redshift. 
}

\item{
 The absolute distributions of disk-dominated and bulge-dominated AGN hosts are similar; both uniformly span the color space at higher masses with similar numbers. There is no evidence for color-mass or morphology trends with respect to the black hole's Eddington ratio.
}

\end{itemize}

\acknowledgments{
  \section*{Acknowledgements} 

We thank the referee for their helpful comments that improved the paper. This work is based on observations made with the NASA/ESA Hubble Space Telescope, obtained from the data archive at the Space Telescope Science Institute. STScI is operated by the Association of Universities for Research in Astronomy, Inc. under NASA contract NAS 5-26555. This work is also based on observations taken by the 3D-HST Treasury Program (GO 12177 and 12328) with NASA/ESA HST, which is operated by the Association of Universities for Research in Astronomy, Inc., under NASA contract NAS5-26555. M.C.P. and C.M.U. are also grateful to Yale University for support, and for summer stipends for S.Y. and M.K. Support for this work was also provided by the National Aeronautics and Space Administration through Einstein Postdoctoral Fellowship Award Number PF5-160143 issued by the Chandra X-ray Observatory Center, which is operated by the Smithsonian Astrophysical Observatory for and on behalf of the National Aeronautics Space Administration under contract NAS8-03060. K.S. gratefully acknowledges support from Swiss National Science Foundation Grant PP00P2\_138979/1.
}

\bibliographystyle{aasjournal}
\bibliography{refs}

\appendix

\section{Comparison to GOODS Morphology in the Literature} \label{app:comp}

\subsection{Van der Wel et al. 2012}

\begin{figure*}[h]
\centering
\includegraphics[width=.7\textwidth]{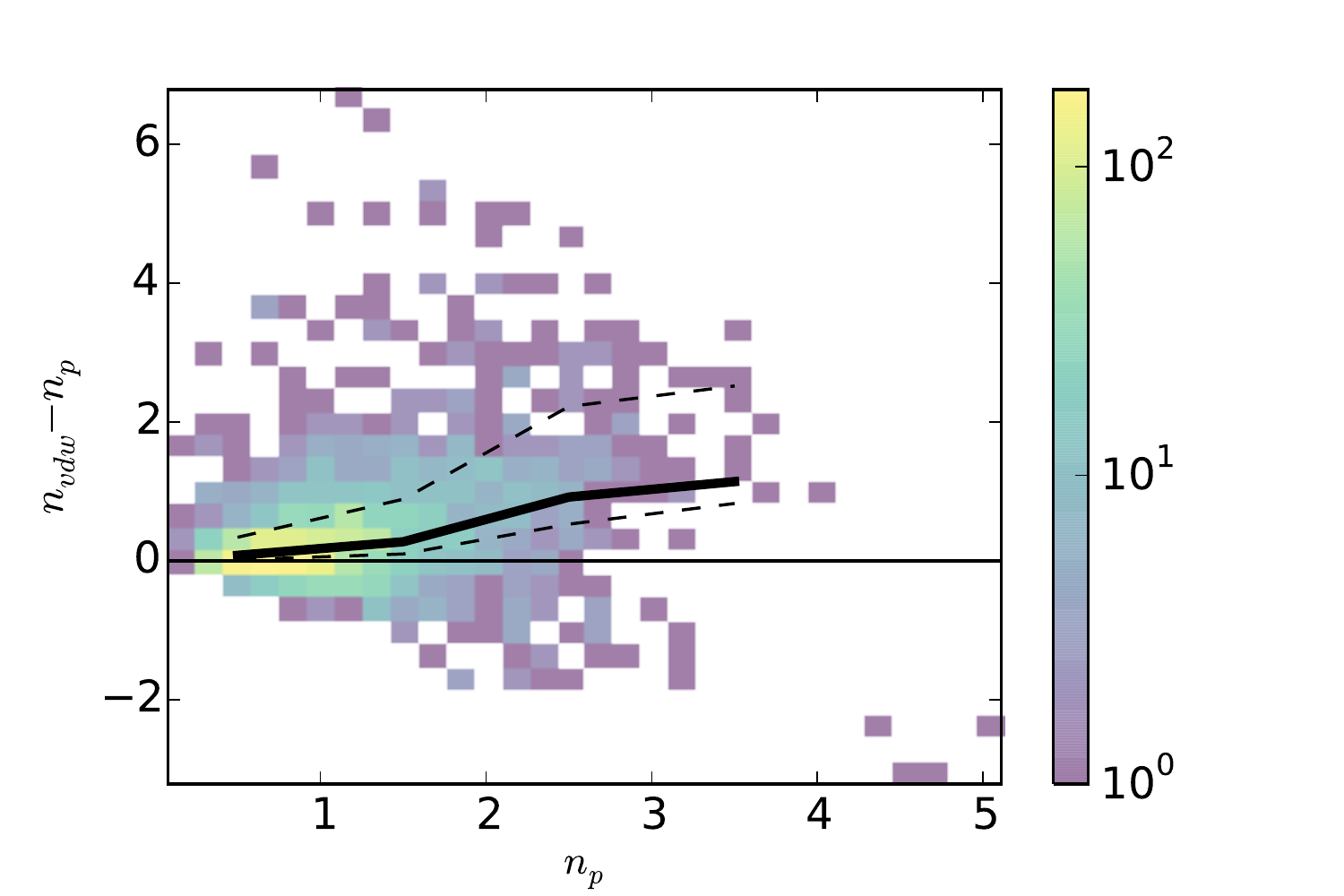}
\caption{Difference between our \sersic\ index values ($n_{p}$; $i$/$z$ band) and those found in van der Wel et al. ($n_{vdw}$; {\it Y} band) as a function of $n_{p}$. The black line shows average values in each bin of $n_{p}$, with the 2D log normal distribution shown in color. We find lower \sersic\ indices, expected since they measure the morphology of the galaxy's older stellar population compared to our analysis. The scatter increases for higher \sersic\ index.}
\label{fig:vdwcomp}
\end{figure*}

\citet{vanderWel12} found structural parameters for $\sim$ 100,000 galaxies in all five CANDELS fields, using GALFIT on YJH images. We found that we obtained systematically lower \sersic\ indices (we will label $n_{p}$) than their values for the Y band ($n_{vdw}$). The difference of the two \sersic\ indices ($n_{vdw}-n_{p}$) is shown in Figure \ref{fig:vdwcomp}. 

The difference can be attributed to the fact that the redder images that \citet{vanderWel12} used measured the morphology of the older stellar population of the galaxy, which tends be more spheroidal than the star forming disk. However we note that had we used their \sersic\ index determinations, $\sim 6 \%$ would be classified differently. The color mass distributions largely look similar.

\subsection{Bundy et al. 2005}

\begin{figure*}[h]
\centering
\includegraphics[width=.95\textwidth]{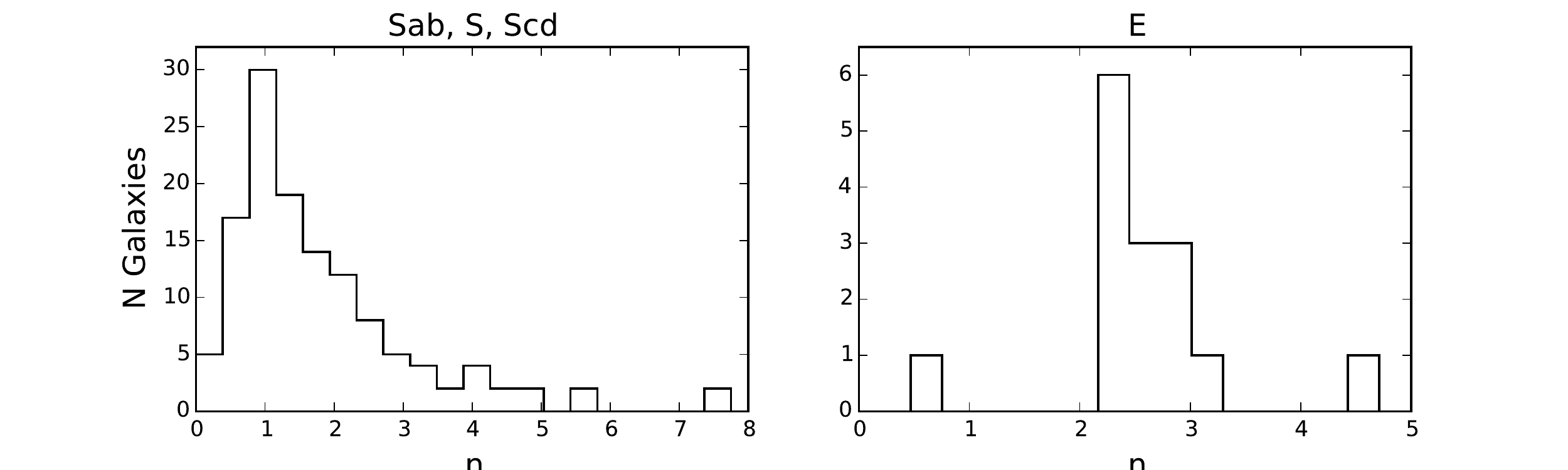}
\caption{Distributions of \sersic\ index for galaxies classified as late-type (left) and early-Type (right) in \citet{Bundy05}. Our classification of disk-dominated galaxies largely agree with galaxies classified as late-type, and classified ellipticals in \citet{Bundy05} mostly agree with our bulge-dominated determination.}
\label{fig:bcomp1}
\end{figure*}

\begin{figure*}[h]
\centering
\includegraphics[width=.7\textwidth]{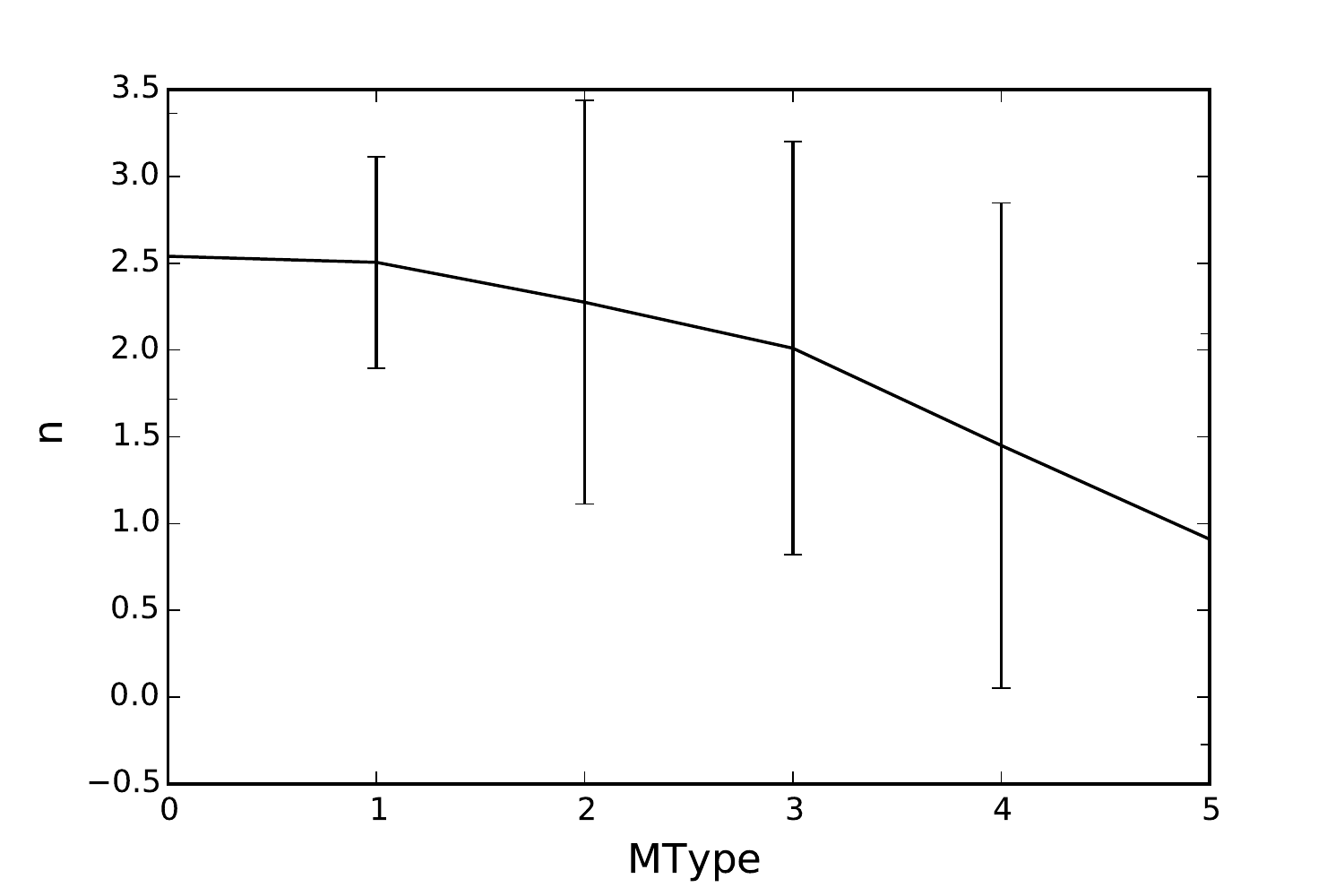}
\caption{The median \sersic\ index for each \citet{Bundy05} morphology type. Larger 'Mtype' corresponds to more spiral, or late-type galaxies, and errorbars represent the standard deviation in each bin. It can be seen while our \sersic\ indices decrease as the Bundy morphology shifts to late-type, as expected, there is a large spread in each bin.}
\label{fig:bcomp2}
\end{figure*}

\citet{Bundy05} assigned morphological types to $z$-band GOODS galaxies ($z_{AB}<22.5$). Figure \ref{fig:bcomp1} shows the distributions of \sersic\ indices for classified spiral galaxies (types Sab, S, and Scd) and for elliptical galaxies (E). The majority of spiral (late-type) galaxies classified in \citet{Bundy05} were also found to have $n<2$ in our analysis. Conversely, elliptical galaxies were mostly measured to have $n>2$. We also plot the median of our \sersic\ index for each morphological type determined in \citet{Bundy05} (which goes from 0 to 5 along the Hubble diagram, corresponding to E, E/S0, S0, Sab, S, Scd, respectively) in Figure  \ref{fig:bcomp2}. While there is a large spread, our \sersic\ indices decrease as the Bundy morphology shifts to late-type, as expected.

\end{document}